\documentclass{emulateapj}

\newcommand{\rmd}{ {\ \mathrm d} }
\renewcommand{\vec}[1]{\mbox{\boldmath $#1$}}

\def\bl{\par\vskip 12pt\noindent}

\shorttitle{Grand Minima and North-South Asymmetry}
\shortauthors{S.\,V.~Olemskoy and L.\,L.~Kitchatinov}

\begin{document}

\title{Grand Minima and North-South Asymmetry of Solar Activity}

\author{S.\,V.~Olemskoy$^1$ and L.\,L.~Kitchatinov$^{1,2}$}
\bl
\affil{$^1${\it Institute for Solar-Terrestrial Physics, Lermontov Str. 126A, Irkutsk 664033, Russia}\\
$^2${\it Pulkovo Astronomical Observatory, St. Petersburg 176140, Russia}}



\begin{abstract}
A solar-type dynamo model in a spherical shell is developed with allowance for random dependence of the poloidal field generation mechanism on time and latitude. The model shows repeatable epochs of strongly decreased amplitude of magnetic cycles similar to the Maunder minimum of solar activity. Random dependence of dynamo-parameters on latitude brakes the equatorial symmetry of generated fields. The model shows the correlation of the occurrence of grand minima with deviations in the dynamo-field from dipolar parity. An increased north-south asymmetry of magnetic activity can, therefore, be an indicator of transitions to grand minima. Qualitative interpretation of this correlation is suggested. Statistics of grand minima in the model are close to the Poisson random process indicating that the onset of a grand minimum is statistically independent of preceding minima.
\end{abstract}

\bl


\section{INTRODUCTION}
Grand minima and North-South asymmetry of solar activity may be mutually related. \citet{RN93} found that almost all sunspots at the end of the Mounder minimum were in the southern hemisphere. The asymmetry reduced in the course of the return of the sun to its \lq normal' activity level so that several decades after the Mounder minimum the asymmetry was no longer pronounced    in sunspot records \citep{A09}. It may be noted also that polar field reversal in the (unusually low) current 24th cycle of solar activity is highly asymmetric \citep{SK13}.

This paper shows that the correlation of the magnetic activity level with its hemispheric asymmetry  naturally emerges in the dynamo theory and the increased asymmetry in epochs of global minima can be reproduced by solar dynamo models.

The currently leading theory of grand minima (and maxima) of solar activity explains them by randomness in dynamo parameters. The relevant literature is extensive. Without attempting a complete referencing, the papers of \citet{H88}, \citet{C92}, and \citet{Oea96} introducing the basic ideas can be quoted. Global magnetic fields are generated by turbulent convection. Therefore, dynamo-parameters experience random fluctuations. Random changes in magnetic field generation mechanisms on time scale of months and years result in irregular variations of magnetic cycle amplitudes on much longer scales of decades and centuries so that the sun spends about 17\% of its life in states of subnormal magnetic activity or grand minima \citep{CK12}. Random fluctuations in dynamo parameters are also essential for reproducing the observed Waldmeier relation \citep{CD00} and polar field reversals \citep{MKS13} by dynamo models.

Randomness of dynamo is naturally allowed for by 3D numerical simulations where dynamically chaotic fluctuations are automatically included. However, in spite of remarkable progress in direct numerical simulations of solar dynamo \citep{GCS10,KMB12,MYK13}, work on the dynamo theory of grand minima was done almost entirely within the mean-field approach. Long-term 3D runs remain too demanding numerically. Mean-field theory has, however, a conceptual difficulty in allowance for randomness of dynamo-parameters. The theory operates with averaged parameters. The averaging procedure is usually understood as ensemble-averaging. Random fluctuations cannot be expected in so-averaged parameters. Ensemble averaging is, however, not the type of averaging of practical use in processing observational data, where temporal or spatial averaging is applied instead. If the averaging is, e.g., over longitude, the resulting axisymmetric parameters still fluctuate. The Babcock-Leighton mechanism of poloidal field generation is related to properties of solar active regions (cf., e.g., \citeauthor{Char10} \citeyear{Char10}). Even in the epochs of activity maxima, several active regions only are simultaneously present on the sun, so that fluctuations of the same order or even exceeding mean values can be expected.

Most work on dynamos with fluctuating parameters concerns fluctuations with time only. Random dependencies on position were disregarded (see, however, \citeauthor{USM09} \citeyear{USM09}). We have to include random dependence on latitude in dynamo parameters in order to account for hemispheric asymmetry induced by the fluctuations. This paper concerns a solar dynamo model with latitude and time dependent fluctuations in the poloidal field generation mechanism.

When the fluctuations are neglected, dynamo-modes can possess a certain equatorial symmetry. The threshold dynamo number for excitation of the fields of dipolar equatorial symmetry in our model is considerably smaller compared to quadrupolar fields. Thus, the model prefers equator anti-symmetric global fields that dominate on the sun \citep{S88}. \citet{CNC04} addressed the parity issue for the Babcock-Leighton type dynamos to conclude that preference for dipolar or quadrupolar parity is defined by whether the characteristic time of diffusion or advection by the meridional flow, respectively, is shorter. Preference for dipolar parity in our model is most probably related to relatively large turbulent diffusion or, more specifically, to relatively low ($\sim 10$) magnetic Reynolds number of the meridional flow.

Latitude-dependent fluctuations in dynamo-parameters violate the equatorial symmetry.
A part of magnetic energy is transmitted to quadrupolar fields that are not supported by the dynamo and thus decay. Accordingly, the model shows anti-correlation between deviations from dipolar parity and amplitudes of magnetic cycles. Substantial hemispheric asymmetry is typically present in the epochs of low magnetic cycles, similar to the observations of \citet{RN93} and \citet{NSRK94}.

\section{DYNAMO MODEL}
Our dynamo model is very close to that of the preceding paper (\citeauthor{KO12} \citeyear{KO12}; KO12 hereafter). The main difference is that now we allow for random fluctuations in dynamo-parameters. Only what is related to these fluctuations will be described in all details. The rest of the model description is nevertheless complete though concise. More details can be found in KO12.

\subsection{Dynamo Equations}
Dynamo equations are formulated for the mean magnetic field $\vec B$ which is assumed to be axisymmetric (longitude-averaged). Detailed discussions of the mean-field electrodynamics and the methods used to derive the mean-field induction equation can be found in literature (cf., e.g., \citet{KR80} or a more recent formulation by \citet{BS05}). Our model employs the simple version of the equation,
\begin{equation}
    \frac{\partial{\vec B}}{\partial t} =
    {\vec\nabla}\times\left[{\vec V}\times{\vec B} - \sqrt{\eta_{_\mathrm{T}}}\ {\vec\nabla}\times\left(
    \sqrt{\eta_{_\mathrm{T}}} \vec{B}\right) + {\vec{\cal A}}\right] ,
    \label{1}
\end{equation}
that, however, includes all the basic effects. The first term in the RHS describes advection of the field $\vec B$ by the mean flow with the velocity $\vec V$. Small-scale motions give rise to the turbulent diffusion with the diffusivity $\eta_{_\mathrm{T}}$.  

Note that the second term in the RHS of Eq.\,(\ref{1}) accounts for not only turbulent diffusion but also for diamagnetic pumping of the field with effective velocity
\begin{equation}
    \vec{U}_\mathrm{dia} = - \frac{1}{2}\vec{\nabla}\eta_{_\mathrm{T}} .
    \label{4}
\end{equation}
Turbulent conducting fluid acts as a diamagnet: large-scale fields are expelled from regions of relatively high intensity of turbulent convection (\citeauthor{KR80} \citeyear{KR80}, chapter 9.5). Diamagnetic pumping is very important for our model. Only with allowance for the pumping was it possible to reproduce basic parameters of the solar cycles (KO12). The pumping concentrates magnetic fields at the base of the convection zone.

The last term in Eq.\,(\ref{1}),
\begin{equation}
  {\vec{\cal A}} = \alpha_0 R_\odot^{-3} \int \alpha({\vec r},{\vec r}')\vec{B}({\vec r}')\ \mathrm{d}^3r' ,
  \label{3}
\end{equation}
accounts for the nonlocal alpha-effect \citep{BRS08}, which in our model represents the Babcock-Leighton mechanism.
The mechanism generates a poloidal magnetic field near the surface from a deep toroidal field. In equation (\ref{3}), $\alpha_0$ is the characteristic value of the alpha-effect and $\alpha ({\vec r},{\vec r}')$ is the dimensionless function, which is not small (of the order of one) if the position $\vec r$ is close to the surface and ${\vec r}'$ is close to the base of the convection zone, but small otherwise.

Toroidal and poloidal parts can be separated in the large-scale flow and magnetic field,
\begin{equation}
    \vec{V} = \vec{e}_\phi r\sin\theta\ \Omega f(r,\theta)  +
    \frac{1}{\rho}\vec{\nabla}\times\left(\vec{e}_\phi\frac{\psi}{r\sin\theta}\right) ,
    \label{5}
\end{equation}
\begin{equation}
    {\vec B} = {\vec e}_\phi B + {\vec\nabla}\times
    \left({\vec e}_\phi\frac{A}{r\sin\theta}\right) ,
    \label{6}
\end{equation}
where the usual spherical coordinates $(r,\theta ,\phi )$ are used, ${\vec e}_\phi$ is the unit vector in azimuthal direction, $\Omega$ is the characteristic value of angular velocity, $f$ is the normalized rotation rate, $\psi$ is the stream function of meridional flow, $B$ is the toroidal field, and $A$ is the poloidal field potential.

Normalized variables are used in dynamo equations. Time is measured in
units of $R_\odot^2/\eta_0$; $\eta_0$ is the characteristic value of
the eddy diffusivity. The magnetic field is normalized to its \lq equipartition'
strength $B_0$ for which nonlinear effects become essential, and the poloidal field potential is measured in units of $\alpha_0B_0R^3_\odot/\eta_0$. The density is normalized to its
surface value $\rho_0$, and the stream function of meridional flow
is measured in units of $\rho_0 R_\odot^2V_0$; $V_0$ is the
amplitude of the surface meridional flow. From now on, the same
notations are kept for the normalized variables as used before for
their dimensional counterparts, except for the fractional radius $x
= r/R_\odot$ and normalized diffusivity $\eta =
\eta_{_\mathrm{T}}/\eta_0$.

The normalized equation for the toroidal field reads
\begin{eqnarray}
    \frac{\partial B}{\partial t} &=&
    \frac{\cal D}{x} \left(\frac{\partial f}{\partial x}\frac{\partial
    A}{\partial\theta} - \frac{\partial f}{\partial\theta}
    \frac{\partial A}{\partial x}\right)
    \nonumber \\
    &+& \frac{R_\mathrm{m}}{x^2\rho}\frac{\partial}{\partial\theta}
    \left(\frac{B}{\sin\theta}
    \frac{\partial\psi}{\partial x}\right)
    - \frac{R_\mathrm{m}}{x\sin\theta}\frac{\partial}{\partial x}
    \left(\frac{B}{\rho x}\frac{\partial\psi}{\partial\theta}
    \right)
    \nonumber \\
    &+&  \frac{\eta}{x^2}\frac{\partial}{\partial\theta}\left(
    \frac{1}{\sin\theta}\frac{\partial(\sin\theta
    B)}{\partial\theta}\right)
    \nonumber \\
    &+& \frac{1}{x}\frac{\partial}{\partial
    x}\left(\sqrt{\eta}\ \frac{\partial(\sqrt{\eta}\ xB)}
    {\partial x}\right) ,
    \label{7}
\end{eqnarray}
where
\begin{equation}
    {\cal D} = \frac{\alpha_0 \Omega R_\odot^3}{\eta_0^2}\
 \label{8}
\end{equation}
is the dynamo number and
\begin{equation}
    R_\mathrm{m} = \frac{V_0 R_\odot}{\eta_0}
    \label{9}
\end{equation}
is the magnetic Reynolds number for the meridional flow. Equation (\ref{7}) describes the toroidal field production by differential rotation, its advection by the meridional flow, diamagnetic pumping, and turbulent diffusion.

The poloidal field equation, which accounts for random fluctuations in the field generation term, is written as
\begin{eqnarray}
    \frac{\partial A}{\partial t} &=&
    x \sin^3\theta \left(\cos\theta + \frac{1}{2}\sigma s(t,\theta)\right)
    \int\limits_{x_\mathrm{i}}^x
    \hat\alpha (x,x') B(x',\theta)\ \mathrm{d} x'
    \nonumber \\
    &+& \frac{R_\mathrm{m}}{\rho x^2 \sin\theta}
    \left(\frac{\partial\psi}{\partial x}
    \frac{\partial A}{\partial\theta} -
    \frac{\partial\psi}{\partial\theta}
    \frac{\partial A}{\partial x}\right)
    \nonumber \\
    &+& \frac{\eta}{x^2}\sin\theta\frac{\partial}{\partial\theta}
    \left(\frac{1}{\sin\theta}\frac{\partial
    A}{\partial\theta}\right) + \sqrt{\eta}\frac{\partial}{\partial
    x} \left(\sqrt{\eta}\frac{\partial A}{\partial x}\right)  ,
    \label{10}
\end{eqnarray}
where $x_\mathrm{i}$ is the radius of the inner boundary. The integration in the first term in the RHS is only in the radius with the upper limit $x$. This qualitatively reflects the fact that generation of poloidal fields at some point $x$ is contributed by the buoyant magnetic loops rising from deeper layers and that the rise velocities are almost vertical. The quantity $\sigma$ in Eq.\,(\ref{10}) is the relative amplitude of fluctuations in the generation term. The mean value of the generation term is antisymmetric about the equator, as it should be for the effect of the Coriolis force. We do not expect, however, that the fluctuations are equator-antisymmetric or vanishing at the equator. Randomness in the generation mechanism is  accounted for by the random function $s$ of time and latitude; the RMS value of $s$ is unity. This function is defined in Section\,2.3.

The Babcock-Leighton mechanism is related to the tilts of bipolar magnetic regions, i.e. to the fact that the regions of magnetic polarity leading in rotational motion are on average closer to the equator than the following polarity. Observations summarised by \citet{H96} and theoretical computations of \citet{DC93} both show that the tilt angles do not increase steadily with latitude but have a maximum. The factor $\sin^3\theta$ in the first term of the RHS of Eq.\,(\ref{10}) reflects this non-monotonous dependence (for the often assumed $\cos\theta$ profile of the alpha-effect, the factor would be $\sin\theta$). Latitudinal distribution of toroidal fields in our model depends, however, little on the latitudinal profile of the alpha-effect (KO12). Advection of the field by the meridional flow is much more significant.

The initial value problem for dynamo equations (\ref{7}) and (\ref{10}) was solved numerically by the grid-point method and explicit time stepping. After several diffusion times, the solution becomes independent of the initial condition. The diamagnetic pumping results in a high concentration of the magnetic fields near the bottom boundary where diffusivity is relatively low. A non-uniform grid in radius with spacing $\Delta x \sim \sqrt{\eta }$ was applied to resolve the fine radial structure. The boundary conditions assume a superconductor beneath the bottom and vertical field (pseudo-vacuum condition) on the top boundary.

\subsection{Model Design}
\begin{figure}[htb]
    \centering
    \includegraphics[width=0.39\textwidth]{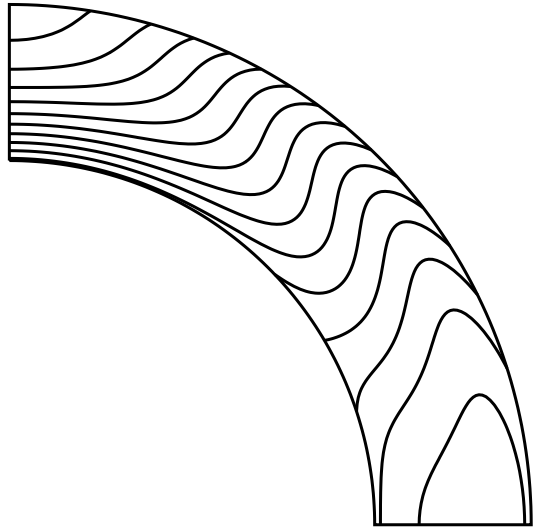}
    \caption{Angular velocity isolines for the differential rotation of our dynamo model.
    }
    \label{f1}
\end{figure}

The bottom boundary in our model is placed at $x_\mathrm{i} = 0.7$, slightly below the bottom of the convection zone at $x = 0.713$ in the sun \citep{BA97}. The differential rotation is specified using the approximation suggested by \citet{BKS00} for the helioseismological rotation law,
\begin{equation}
    f(x,\theta) = \frac{1}{461}\sum\limits_{m=0}^{2}
    \cos\left( 2m\left(\frac{\pi}{2}
    - \theta\right)\right) \sum\limits_{n=0}^{4} C_{nm}x^n .
    \label{11}
\end{equation}
Numerical values of the coefficients $C_{nm}$ are given in \citeauthor{BKS00} The rotation law is shown in Fig.\,\ref{f1}.

\begin{figure}[htb]
    \resizebox{\hsize}{!}{\includegraphics{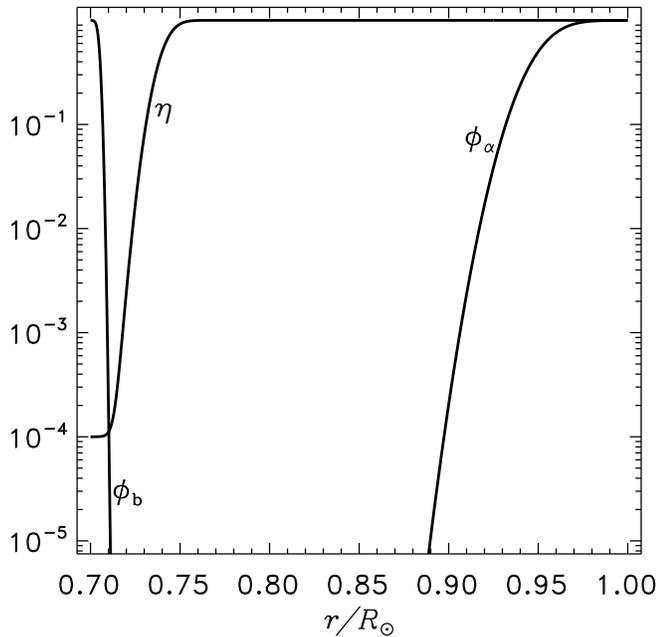}}
    \caption{Profiles of the normalized diffusivity and kernel functions of Eq.\,(\ref{12})
        used in the dynamo model.
    }
    \label{f2}
\end{figure}

The function $\hat{\alpha}(x,x')$ in the generation term of Eq.\,(\ref{10}) was specified as follows,
\begin{eqnarray}
    \hat\alpha (x,x') &=& \frac{\phi_\mathrm{b}(x')\phi_\alpha (x)} {1 + B^2(x',\theta)}
    ,
    \nonumber \\
    \phi_\mathrm{b}(x') &=& \frac{1}{2}\left( 1 -
    \mathrm{erf}\left( (x' - x_\mathrm{b})/h_\mathrm{b}\right)\right) ,
    \nonumber \\
    \phi_\alpha (x) &=& \frac{1}{2}\left( 1 +
    \mathrm{erf}\left( (x - x_\alpha)/h_\alpha\right)\right),
    \label{12}
\end{eqnarray}
where $B^2$ in the denominator introduces the only nonlinearity of the model as the magnetic quenching of the generation effect, and $\mathrm{erf}$ is the error function. The function $\phi_\mathrm{b}(x')$ defines the region near the bottom whose toroidal fields take part in the generation of poloidal fields near the surface. The function $\phi_\alpha (x)$ defines the near-surface region where the generation takes place. The parameters $h_\mathrm{b}$ and $h_\alpha$ are the thicknesses of the near-bottom and near-surface layers, respectively. We always take $x_\mathrm{b} = x_\mathrm{i} + 2.5h_\mathrm{b}$ and $x_\alpha = 1 - 2.5h_\alpha$ to ensure smoothness of the functions (\ref{11}) in the simulation domain. Computations of this paper were performed with $h_\alpha = 0.02$ and $h_\mathrm{b} = 0.002$. Figure~\ref{f2} shows the kernel functions of Eq.\,(\ref{12}) together with the diffusivity profile used in the computations.

The turbulent diffusivity varies mildly in the bulk of the convection zone but decreases sharply with depth near its base. The diffusivity profile of our model mimics this behavior
\begin{equation}
    \eta (x) = \eta_\mathrm{in} + \frac{1}{2}(1 - \eta_\mathrm{in})
    \left( 1 + \mathrm{erf}\left(\frac{x -
    x_\eta}{h_\eta}\right)\right) .
    \label{13}
\end{equation}
The $\eta_\mathrm{in}$ in this equation is the ratio of radiation interior-to-convection zone diffusivities. The diffusivity drops from convection to radiation zone in the sun by about nine  orders of magnitude. Computations were performed for the smallest value of $\eta_\mathrm{in} = 10^{-4}$ accessible in the numerical model. Figure~\ref{f2} shows the diffusivity profile for the parameters $x_\eta = 0.74$ and $h_\eta = 0.02$ used in the computations. The thin bottom layer of small diffusivity is very important for the model. Almost all toroidal magnetic flux is contained in this layer due to the downward diamagnetic pumping of Eq.~(\ref{4}).

\begin{figure}[htb]
    \centering
    \includegraphics[width=0.45\textwidth]{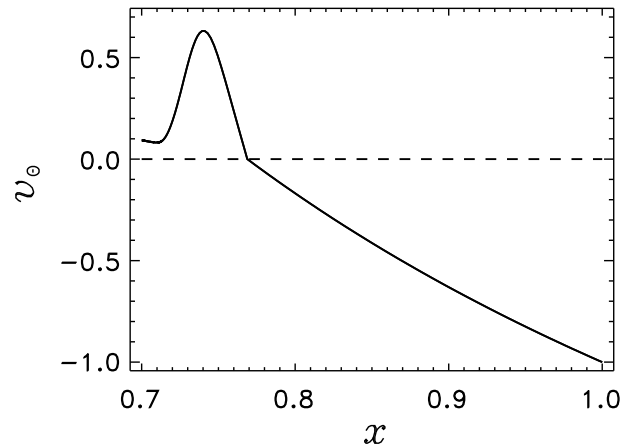}
    \caption{Radial profile of normalised meridional velocity used in the dynamo model.
    }
    \label{meri}
\end{figure}

The meridional flow of the model has one circulation cell per hemisphere with a poleward flow on the top and a return equator-ward motion near the bottom.
Poleward surface flow of the order of 10\,m/s is observed on the sun \citep{KHH93}. Helioseismoilogy finds this flow to persist up to a depth of about 12 Mm \citep{ZK04}. Measurements of the deep near-bottom flow give controversial results. Theoretical models of global solar circulation predict that the return flow at the base of the convection zone is not  small compared to the surface \citep{RKA05, KO11}. The flow, however, decreases rapidly with depth beneath the base \citep{GM04}. All these characteristics are qualitatively reflected by the following prescription for the stream-function of Eq.\,(\ref{5})
\begin{eqnarray}
       \psi &=&- \cos\theta\ \sin^2\theta\ \Phi(x) ,
    \nonumber  \\
    \Phi(x) &=& \left\{\begin{array}{ll}
            \frac{1}{1 - x_\mathrm{s}} \int\limits_x^1 \rho
            (x')\eta^p(x')(x'-x_\mathrm{s}) \rmd x' & \mbox{for $x
            \geq x_\mathrm{s}$} \\
            C \int\limits_{x_\mathrm{i}}^x \rho(x') \eta^p(x')
            (x_\mathrm{s} - x') \rmd x' & \mbox{for $x
            \leq x_\mathrm{s} .$}
            \end{array}
            \right.
            \label{mf}
\end{eqnarray}
used in the dynamo model. In this equation, $x_\mathrm{s}$ is the radius of the stagnation point where the meridional velocity changes sign, and $C$ is a parameter whose value is adjusted to ensure continuity of the stream function at the stagnation point. An adiabatic profile for ideal gas was used for the normalised density,
\begin{equation}
    \rho(x) = \left( 1 + C_\rho\left(\frac{1}{x}
    -1\right)\right)^{3/2} ,
    \label{rho}
\end{equation}
where $C_\rho = 10^3$. Computations in this paper were performed with $p =1/3$ and $x_\mathrm{s} = 0.77$ in Eq.\,(\ref{mf}). The depth profile of the meridional flow is shown in Fig.\,\ref{meri}.

The magnetic Reynolds number (\ref{9}) was taken to be $\mathrm{Rm} = 10$. This implies the diffusivity value $\eta_0 \simeq  10^{13}$\,cm$^2$s$^{-1}$ in the bulk of the convection zone favored by the mixing-length estimations. It may be noted that mean-field models with smaller values of the eddy viscosity and thermal diffusivity are hydrodynamically unstable \citep{Tea94}. The smaller values, therefore, cannot be a correct parameterization for solar convection. Direct numerical simulations of \citet{YBR03} suggest that the eddy magnetic diffusivity is of the same order of magnitude as the eddy viscosity.

Computations were performed for the dynamo number of Eq.~(\ref{8}) equal ${\cal D} =3.2\times 10^4$, slightly above the critical value for excitation of dipolar modes (${\cal D}^\mathrm{d}_\mathrm{c} = 3.01\times 10^4$) but below the threshold value for quadrupolar fields (${\cal D}^\mathrm{q}_\mathrm{c} = 3.79\times 10^4$). Slight difference in the dynamo numbers with KO12 is explained by the difference in the meridional flow.

\subsection{Smoothly Varying Random Function of Time and Latitude}
Randomness of the poloidal field generation mechanism is represented in the dynamo equation (\ref{10}) by the random function $s(t,\theta )$. We use a method very close to that of \citet{R05} to model this function.

Random dependence on time only can be modelled by solving numerically the equation
\begin{equation}
    \frac{\mathrm{d} s}{\mathrm{d} t} = -\frac{s}{\tau} + \frac{g}{\tau} ,
    \label{14}
\end{equation}
where $\tau$ is the characteristic time scale of fluctuations, and $g$ is the random number renovated each time step of a numerical computation. The resulting random dependence on time can be characterised by the correlation function
\begin{equation}
    \varphi (t) = \langle s(t_0 + t)s(t_0)\rangle .
    \label{15}
\end{equation}
Applying the random forcing $g = \hat{g}\sqrt{2\tau/\Delta t}$ ($\hat{g}$ is the random number with Gaussian distribution, zero mean, RMS value equal one, and renovated each time step $\Delta t$ independently of its former value), we get
\begin{equation}
    \varphi (t) = \mathrm{exp}\left( -\mid t\mid /\tau \right)
    \label{16}
\end{equation}
for the case of a small time step $\Delta t \ll \tau$. Note that the forcing $g$ is short correlated in time but the characteristic time $\tau$ of the correlation (\ref{16}) is much longer. The correlation (\ref{16}), however, is singular at $t=0$. The function $s(t)$ is dominated by the latest realization of the forcing $g$. In order to make the function $s(t)$ more smooth, we apply the two-steps scheme
\begin{eqnarray}
    \frac{\partial s}{\partial t} &=& -\frac{s}{\tau} + \frac{s_1}{\tau} ,
    \nonumber \\
    \frac{\partial s_1}{\partial t} &=& -\frac{s_1}{\tau} + \frac{g}{\tau} .
    \label{17}
\end{eqnarray}
Equations (\ref{17}) with the forcing $g = \hat{g}\sqrt{4\tau /\Delta t}$ ($\hat{g}$ is the same Gaussian random number as before) leads to the correlation
\begin{equation}
    \varphi (t) = \left(\frac{\mid t \mid }{\tau} + 1\right)
    \mathrm{exp}\left(-\mid t\mid /\tau \right) ,
    \label{18}
\end{equation}
which is now smooth at $t = 0$. It was confirmed by computations that the numerical solution of Eq.\,(\ref{17}) with forcing $g$ renovated each time step reproduces the correlation function (\ref{18}).

To involve random dependence not on time only but on the latitude as well, we use the forcing
\begin{equation}
    g =  (2\pi)^{1/4} \sqrt{\frac{4\tau}{\Delta t \delta\theta}}
    \mathrm{exp}\left(-\left(\frac{\theta-\theta_0}{\delta\theta}\right)^2\right) \hat{g}.
    \label{19}
\end{equation}
In this equation, $\delta\theta$ is the characteristic scale of fluctuations in latitude, and $\theta_0$ is the random co-latitude uniformly distributed in the range $[ 0,\pi ]$ between the north and south poles. The random value of $\theta_0$ was also renovated each time step.

\begin{figure*}[htb]
    \centering
    \includegraphics[width=0.77\textwidth]{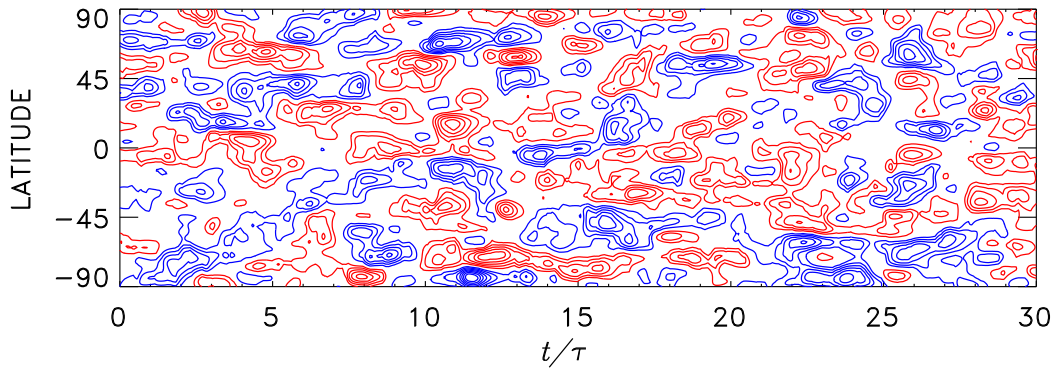}
    \caption{Isolines of random function $s(t,\theta )$ on the time-latitude plane for a typical realisation of the function in our computations. Lines of different style show levels of opposite sign. The RMS amplitude of the fluctuating function equals one.
    }
    \label{f3}
\end{figure*}

Equations (\ref{17}) with forcing (\ref{19}) are solved numerically together with the dynamo equations to realise the random dependence of the poloidal field generation on time and latitude. The random function $s(t,\theta )$ has RMS value close to one everywhere except for the near-polar regions of extent $\delta\theta$. The latitudinal scale of fluctuations in our computations $\delta\theta = 0.1\ (\simeq 5.7^\circ )$, is close to the latitudinal extent of solar active regions.  Characteristic realisation of the random function is shown in  Fig.\,\ref{f3}. The time scale $\tau$ was taken equal to the period of solar rotation. The amplitude of the fluctuations in equation (\ref{10}) $\sigma = 2.7$ was specified after estimations of the fluctuations of the Babcock-Leighton mechanism using sunspot data \citep{OCK13}.

\section{RESULTS AND DISCUSSION}
\begin{figure}[htb]
    \resizebox{\hsize}{!}{\includegraphics{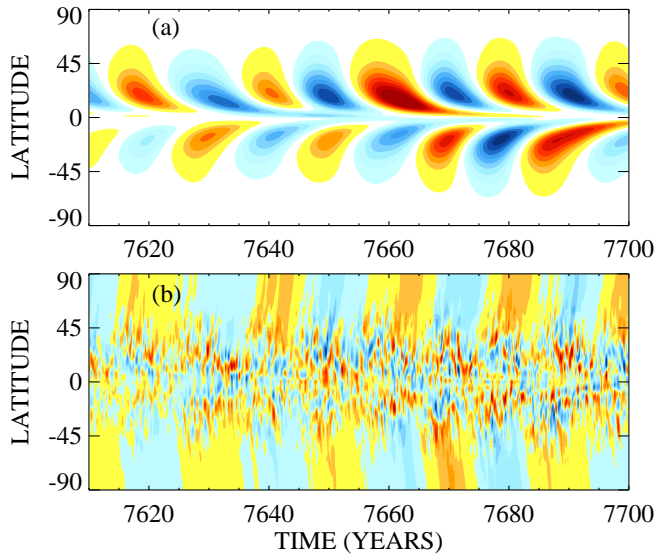}}
    \caption{Time-latitude diagrams of the near-bottom toroidal magnetic field (a) and the surface radial field (b).
    }
    \label{f4}
\end{figure}

Random dependence on latitude in the magnetic field generation mechanism prevents the appearance of magnetic fields of pure equatorial symmetry. The fields are superposition of equator-symmetric or quadrupolar, $B^\mathrm{q}(\theta ) = (B(\theta ) + B(\pi - \theta ))/2$, and equator-antisymmetric or dipolar, $B^\mathrm{d}(\theta ) = (B(\theta ) - B(\pi - \theta ))/2$, components, $B = B^\mathrm{q} + B^\mathrm{d}$. Similar to the field observed on the sun, the dipolar component dominates when the amplitude of magnetic cycles is not too low \citep{S88}, i.e., away from the grand minima. Figure~\ref{f4} shows the time-latitude diagram for such a span of \lq normal' magnetic cycles. We use the diffusivity value $\eta_0 = 10^{13}$\,cm$^2$/s to convert normalised time of numerical computations into physical time of this Figure. With this value, the averaged period of magnetic cycles of 9.7 yr is not far from the observed 11-year period.
The equatorial drift of the toroidal field is a consequence of the field transport by the meridional flow while the polar branches in the poloidal field diagram are caused mainly by the field diffusion from mid latitudes where the alpha-effect of our model has a maximum  (KO12).

The near-bottom toroidal field in Fig.\,\ref{f4} varies with time much more smoothly than the surface poloidal field. This is because the random fluctuations in the generation mechanism are directly affecting the poloidal filed. The fine latitudinal structure is largely smoothed-out in the course of the poloidal field diffusion to the base of the convection zone where the toroidal field is generated by differential rotation.

\begin{figure}[htb]
    \resizebox{\hsize}{!}{\includegraphics{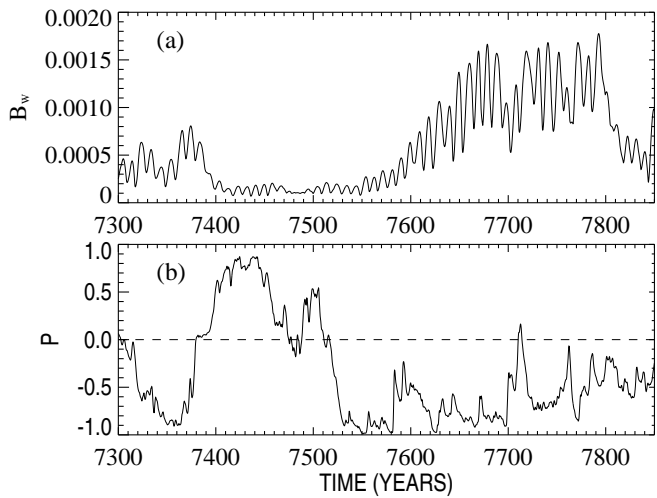}}
    \caption{Amplitude of the simulated magnetic field of Eq.\,(\ref{21}) (a) and the parity index $P$ (b) as a function of time. Note the increase in the parity index at the beginning of \lq grand minimum' at about 7400 yrs.
    }
    \label{f5}
\end{figure}

The total (volume-integrated) magnetic energy can also be represented as the sum of the energies of dipolar and quadrupolar components of the magnetic field, $E = E^\mathrm{q} + E^\mathrm{d}$. We use the standard parity index,
\begin{equation}
    P = \left(E^\mathrm{q} - E^\mathrm{d}\right)/E ,
    \label{20}
\end{equation}
to characterise equatorial symmetry of the simulated fields. The time-dependent index is normally negative indicating predominance of the dipolar component in the generated field. The mean value over the entire simulation time of about 11000 years is $\overline{P} = -0.37$. However, the parity index usually increases when magnetic energy decreases in the course of a transition to a grand minimum. A characteristic example is shown in Fig.\,\ref{f5} where the parity index (\ref{20}) is shown together with the unsigned magnetic flux
\begin{equation}
     B_{_\mathrm{W}} =
    \int\limits_{x_\mathrm{i}}^1\!\!\!\int\limits_0^\pi\! \sin\theta\,x\,\phi_\mathrm{b}(x)\,|B(x,\theta)|\,\mathrm{d}x\,\mathrm{d}\theta
    \label{21}
\end{equation}
in the near-bottom region. This quantity is the latitude-integrated contribution of the near-bottom toroidal field into the terms of Eqs.\,(\ref{10}) and (\ref{12}) responsible for the generation of the near-surface poloidal field. As the generation terms represent the Babcock-Leighton mechanism in our model, the quantity (\ref{21}) can be considered as proxy of  the sunspot number.

\begin{figure}[htb]
    \resizebox{\hsize}{!}{\includegraphics{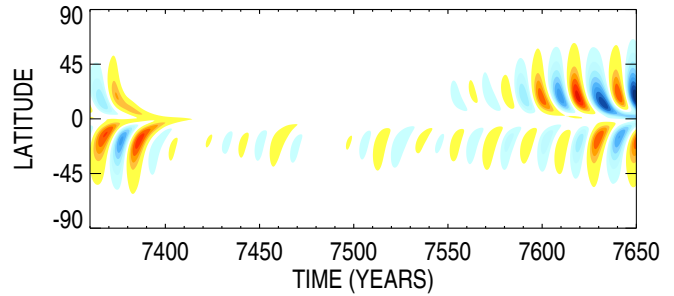}}
    \caption{Butterfly diagram of the near-bottom magnetic field for the period of strongly reduced amplitude of magnetic cycles. The toroidal field is almost entirely in the southern hemisphere in the middle of this period.
    }
    \label{f6}
\end{figure}

On average, the parity index increases when $B_{_\mathrm{W}}$ decreases. The correlation of the deviations of the parity index and the $B_{_\mathrm{W}}$-parameter from their mean values equals -0.30. The negative correlation can be interpreted as follows. The model excites preferentially dipolar fields. Latitude-dependent random fluctuations in the alpha-effect transmit a part of the magnetic energy to quadrupolar modes. The quadrupolar modes are subcritical for dynamo excitation and decay. Deviations from dipolar parity thus reduce the magnetic energy. Large deviations can even promote transitions to grand minima. If this explanation is correct, deviations of the global magnetic field of the sun from an equator-antisymmetric state can be used as an indicator of a decrease in magnetic activity.

Figure\,\ref{f6} shows the butterfly diagram of the near-bottom toroidal field for a simulated grand minimum. The north-south asymmetry is large in the middle of the minimum epoch similar to the butterfly diagram of sunspots observed at the end of the Maunder minimum \citep{RN93}.

\begin{figure*}[htb]
    \centering
    \includegraphics[width=0.77\textwidth]{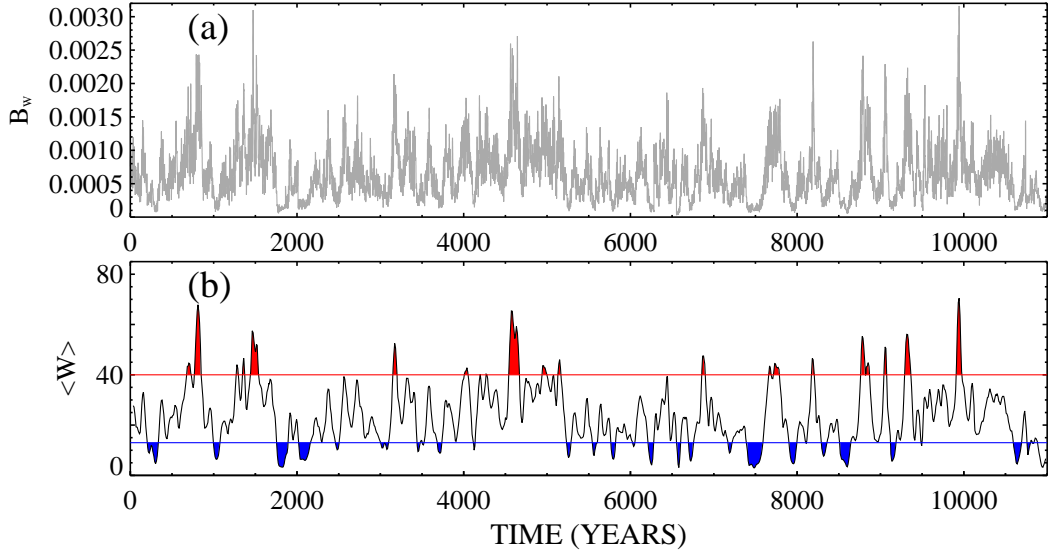}
    \caption{ Dependence of the $B_{_\mathrm{W}}$-parameter (\ref{21}) on time in the long-term dynamo simulation (a). Smoothed amplitude of magnetic cycles (b). The lower and upper horizontal lines show the levels of transition to global minima and global maxima respectively. The regions of the global minima/maxima are shaded.
    }
    \label{f7}
\end{figure*}

There is no reliable recipe for transforming magnetic fields of dynamo models into sunspot number. We use the simplest linear relation between the $B_{_\mathrm{W}}$-parameter of Eq.\,(\ref{21}) and the \lq modelled' sunspot number $W$
\begin{equation}
    W = 33500 \times B_{_\mathrm{W}} ,
    \label{22}
\end{equation}
where the proportionality factor was tuned so that the maximum $W$ in our computations equals the maximum sunspot number reconstructed by \citet{USK07} from radiocarbon data. To compare with the data, we applied the same smoothing procedure to our computations as used by \citeauthor{USK07}
First, amplitudes of magnetic cycles were estimated as maxima of the running annual mean of the $W$-parameter (\ref{22}) for distinct cycles of the dynamo model. Next, the smoothed amplitude $\langle W\rangle$ was estimated by averaging over five neighboring cycles using the Gleisberg's filter 1-2-2-2-1. The time series of the resulting $\langle W\rangle$ values is shown in Fig.\,\ref{f7} together with the original $B_{_\mathrm{W}}$-parameter of Eq.\,(\ref{21}). Two horizontal lines in the bottom panel of this Figure show the levels of transition to grand minima (below the lower line) and maxima (above the upper line). The levels are defined by the conditions that the system spends 17\% of the time in states of grand minima and 9\% - in global maxima \citep{USK07}.

\begin{figure}[htb]
    \begin{center}
    \includegraphics[width=0.23\textwidth]{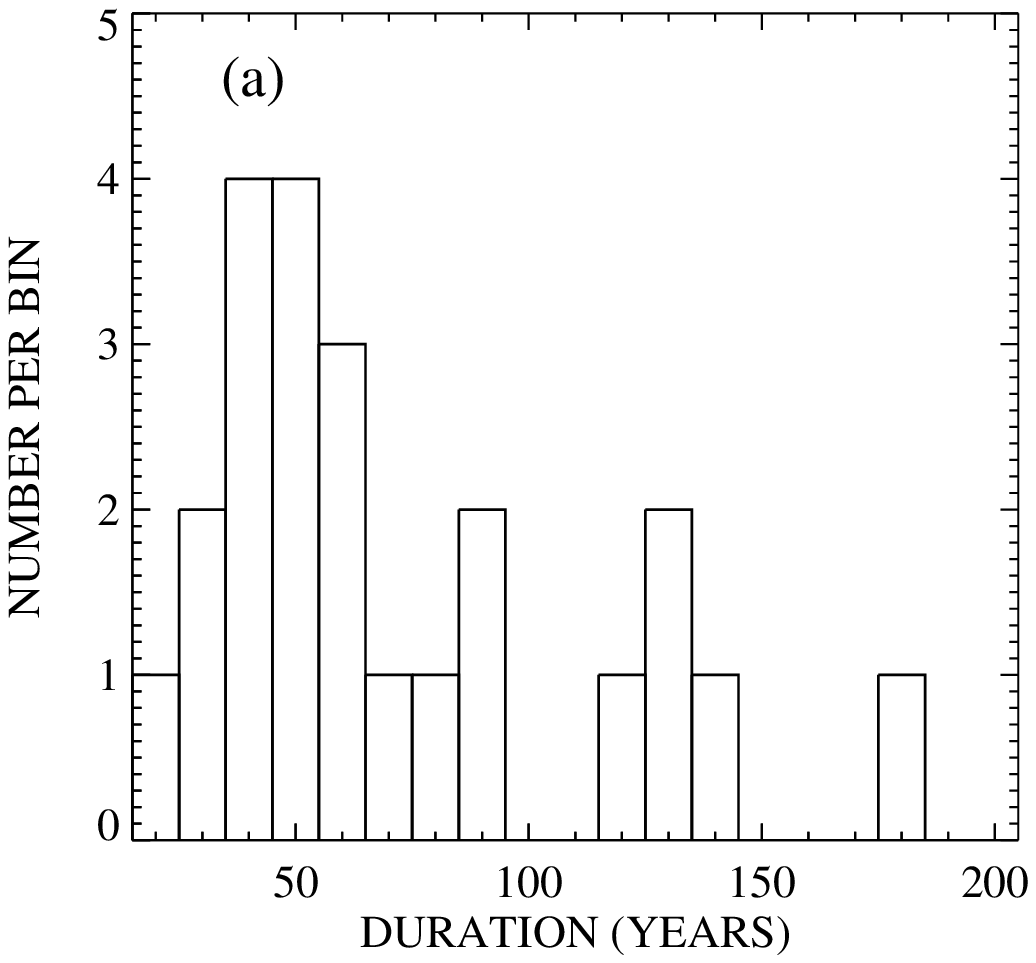}
    \hspace{0.1truecm}
    \includegraphics[width=0.23\textwidth,height= 3.9 cm]{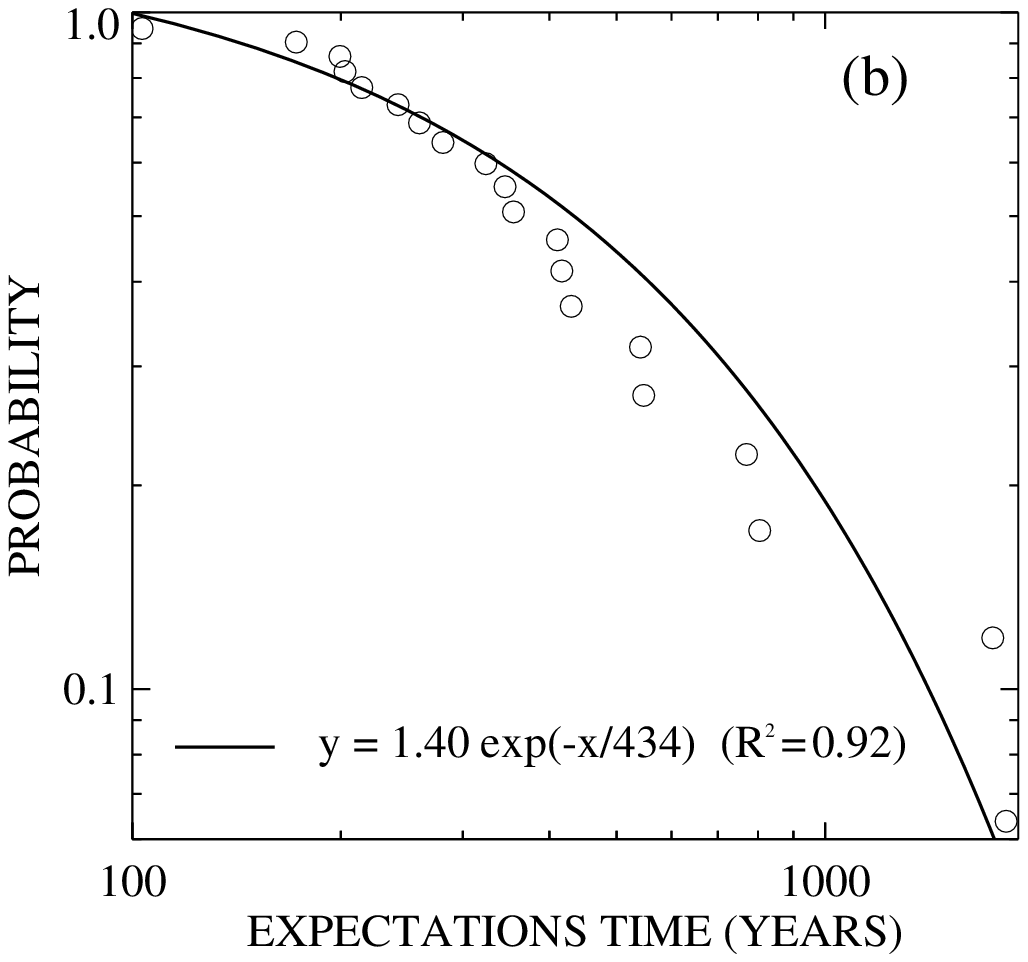}
    \end{center}
    \caption{ (a) Histogram of durations of grand minima in the dynamo model.  (b) Distribution of expectation time of grand minima. The full line shows the exponential fit.
    }
    \label{f8}
\end{figure}

Statistics of grand minima is illustrated by Fig.\,\ref{f8}. The histogram of durations of grand minima shows two groups of relatively short (30-90 years duration) and relatively long ($>$110 years) events. The two groups can be identified with the Mauder-type and Sp\"orer-type minima revealed in the radiocarbon data \citep{SB89,USK07}.

Figure\,\ref{f8}(b) shows the distribution of expectation times of grand minima, i.e., the probability of expecting the next global minimum longer than a given time as a function of this time. The probability ($P$) can be roughly approximated by the exponential law $P = \mathrm{exp}(- t/T)$ corresponding to the Poisson random process. The exponential fit suggests that occurrence of each next minimum is statistically independent of the previous one. Magnetic fields are evolved continuously with time in the dynamo process. Dynamo, therefore, has a finite memory. The memory time would be indefinitely long if random fluctuations were absent. The fluctuations, however, destroy the memory to reduce its characteristic time, probably, to the diffusion time $R^2_\odot /\eta_0$ which is of the order of the magnetic cycle period in our model.
Recently \citet{YNM08} explored and clarified the memory issue for the solar dynamo. They showed that in the diffusion-dominated regime, to which our model belongs, the memory extends to about one solar cycle only. Downward diamagnetic pumping shortens somewhat the memory still further \citep{KN12}.
Expectation times of the global minima are typically much longer than the magnetic cycle period. The subsequent minima, therefore, occur almost independently, as the radiocarbon data also suggest \citep{USK07}.

The results of only one but representative run have been discussed so far. Several other runs for varying dynamo number were performed. The global minima become shorter and eventually disappear as the number increases. We believe that the slightly supercritical number ${\cal D} = 3.2\times 10^4$ is a reasonable choice for the sun. Rotation of solar-type stars is not decelerated beyond a maximum rotation period depending on spectral type \citep{R84}. The maximum period probably corresponds to the rotation rate where global dynamos and angular momentum loss for stellar wind cease. The solar rotation period is only slightly shorter than the maximum period observed for G2 dwarfs \citep{R84}. Only old slowly rotating stars show global minima of magnetic activity \citep{W04}.

\section{SUMMARY}
The model of this paper differs from previous dynamo-models with fluctuating dynamo-parameters by allowance for random dependence of the fluctuations on latitude. Amplitude of the fluctuations in the poloidal field generation mechanism was prescribed after its estimation from sunspot statistics. When the random fluctuations are neglected, the model describes an oscillating global field of dipolar parity similar to the fields of recent solar cycles. Latitude-dependence of the fluctuations violates equatorial symmetry of the dynamo-field. When the deviations from dipolar parity are relatively large, the model typically shows decreased  amplitude of magnetic cycles. The model, therefore, predicts the increased north-south asymmetry of magnetic activity to be a characteristic feature of grand minima. The high correlation of decreased amplitude of magnetic cycles with deviations from dipolar parity in the dynamo model is explained by the dynamo being subcritical to excitation of quadrupolar fields. The explanation implies that the rotation rate of the sun is not much faster than the critical rate for the onset of a global dynamo.

The statistical distribution of expectation times of the modelled grand minima is closely approximated by exponential law. This indicates that the occurrence of grand minima is close to the Poisson-type random process with each subsequent minimum statistically independent on preceding minima. This is similar to the conclusion of \citet{USK07} based on radiocarbon data on global minima occurrence over the Holocene. Randomness in the poloidal field generation mechanism reduces the memory time of the dynamo process, which becomes shorter than the characteristic expectation time of grand minima.

\acknowledgments
The authors are thankful to the Russian Foundation for Basic Research
(projects 12-02-92691\_Ind, 13-02-00277, 12-02-33110-mol\_a\_ved) and to the Ministry of Education and Science of the
Russian Federation (contract 8407) for their support.
\bibliographystyle{apj}

\end{document}